\documentclass[12pt]{article}
\usepackage[centertags]{amsmath}
\usepackage{amssymb}
\usepackage{amsthm}
\usepackage{enumerate}
\usepackage{url}
\usepackage{graphicx}

\usepackage{graphicx}
\usepackage[all, knot]{xy}
\usepackage{amsfonts,amssymb}
\usepackage{latexsym}
\usepackage{amsmath}
\usepackage[all]{xy}
\usepackage{stmaryrd}

\usepackage[english]{babel}

\usepackage{color}
\usepackage[normalem]{ulem}

\makeatletter
  \def\tagform@#1{\maketag@@@{(#1)\@@italiccorr}}
\makeatother

\newtheorem{Theorem}{Theorem}

\newcommand {\RR}{\mbox{${\mathbf{R} }$}}

\newcommand {\D}{\mbox{${\Delta \! \! \! \! \Delta }$}}

\newcommand{\calB}{\cal B}
\newcommand{\R}{\mathbb{R}}
\newcommand{\LL}{\mathbb{L}}
\newcommand{\MM}{\mathbb{M}}
\newcommand{\NN}{\mathbb{N}}
\newcommand{\GG}{\mathbb{G}}
\newcommand{\II}{\mathbb{I}}
\newcommand{\calG}{\mathcal{G}}
\newcommand{\Cinf}{\mbox{$C^{\infty}$}}

\begin{document}

\title{Synthetic Approach to the Singularity Problem}
\author{Michael Heller \\
Copernicus Center for Interdisciplinary Studies\\
Cracow, Poland\\
\and 
Jerzy Kr\'ol \\
Institute of Physics, University of Silesia, \\
ul. Uniwersytecka 
4, 40-007 Katowice, Poland}

\date{\today}
\maketitle

\begin{abstract}
We try to convince the reader that the categorical version of differential 
geometry, called Synthetic Differential Geometry (SDG), offers valuable tools 
which can be applied to work with some unsolved problems of general 
relativity. We do this with respect to the space-time singularity problem. The 
essential difference between the usual differential geometry and SDG is that 
the latter enriches the real line by introducing infinitesimal of various 
kinds. Owing to this geometry acquires a tool to penetrate ``infinitesimally 
small'' parts of a given manifold. However, to make use of this tool we must 
switch from the category of sets to some other suitable category. We try two 
topoi: the topos $\calG $ of germ determined ideals and the so-called Basel 
topos $\calB $. The category of manifolds is a subcategory of both of them. In 
$\calG $, we construct a simple model of a contracting sphere. As the sphere 
shrinks, its curvature increases, but when the radius of the sphere reaches 
infinitesimal values, the curvature becomes infinitesimal and the singularity 
is avoided. The topos $\calB $, unlike the topos $\cal G $, has invertible 
infinitesimal and infinitely large nonstandard natural numbers. This allows us 
to see what happens when a function ``goes through a singularity''. When 
changing from the category of sets to another topos, one must be ready to 
switch 
from classical logic to intuitionistic logic. This is a radical step, 
but the logic of the universe is not obliged to conform to the logic of our 
brains.
\end{abstract}

\section{Introduction}
One of still pending problems of general relativity is the problem of 
space-time singularities. Although there is a common belief that this problem 
will be resolved by the quantum theory of gravity, when finally discovered, so 
far it remains only a belief. And it cannot be excluded that, \textit{vice 
versa}, digging into the singularity problem could help finding the correct 
quantum gravity theory. After all, both these problems are concerned with 
space-time on the smallest possible scale. One should also not forget that the 
singularity problem is highly interesting from the mathematical point of view, 
and it constitutes a good testing field for new mathematical tools.

One of such tools, only recently applied to general relativity (see, for 
instance \cite{Crane,Grin96,GutsGrin96,GutsZvy,Heunen,Kosteckith}), is 
category theory. It is obvious that general relativity in its standard 
formulation is done in the category of sets and functions between sets (SET 
category), just as all macroscopic physical theories (usually without 
explicitly specifying this assumption), and most often in its subcategory of 
smooth manifolds and smooth maps between them ($\MM $ category). However, when 
we approach a level of ``very small'' quantities -- as, for example, in our 
search for quantum gravity and in the singularity problem -- the situation can 
drastically change. Topos formulation of quantum mechanics (see, 
\cite{AC,DoringIsham}) can serve as a motivation for applying similar approach 
to some unsolved problems of general relativity. Category theorists have 
elaborated a categorical version of differential geometry, the so-called 
Synthetic Differential Geometry (SDG) (see, \cite{Kock09}) which almost 
exactly parallels the usual differential geometry employed in relativistic 
calculations. The essential difference consists in the fact that in SDG 
infinitesimals appear which substantially enrich the usual real line. Owing to 
this fact geometry acquires a tool to penetrate infinitesimally small portions 
of a given manifold (a manifold's ``germs'', to be defined below) which in the 
usual approach are invisible (in SET they simply do not exist). This creates 
an invaluable opportunity for physical applications. However, to make use of 
this opportunity we must change from SET to a suitable category. And our 
working hypothesis is that the fundamental level (below Planck's threshold) is 
structured by some other (than SET) category. Of course, some correspondence 
between SET and this category should exist. The minimum requirement is that 
the manifold category $\MM $ must be a subcategory of this new category.  More 
than one choices are possible, and in the following we shall experiment with 
at least two topoi.

In the present paper, we apply the above strategy to the singularity problem 
in general relativity. We aim at blazing a trail rather than directly attack 
the final solution. The paper is addressed to relativity professionals who 
might not be in deep acquainted with category theory. This is why the 
categorical material is presented in as soft manner as possible (only section 
6 is more technical than descriptive). In section 2, for the sake of 
completeness and eventual category theorist readers, we briefly sketch the 
singularity problem in general relativity. In section 3, we introduce 
infinitesimals and show, on the one hand, how they change the perspective of 
doing differential geometry and, on the other, how they enforce a modification 
of logic. To make this new perspective to work, we must put the entire problem 
into the structural environment of a concrete category. We do this in section 
4. A few possibilities are open. Our choice, mainly for simplicity reasons, is 
the topos $\calG $ of germ determined ideals. It is generated by a subcategory 
$\GG$ of the 
category $\LL $ of loci, which we also define and briefly discuss. The 
category $\MM $ of manifolds sits fully and faithfully in $\calG $. Section 4 
is based on our earlier result \cite{MHJK16} that on any infinitesimal 
neighbourhood the components of the Riemann curvature tensor are infinitesimal 
(if not zero). We illustrate this with the help of a simple model of a 
contracting sphere. If we look at the contraction process ``from outside'' (in 
topos SET), as the sphere shrinks, its curvature increases and diverges at 
zero. However, from the ``inside perspective'' (in topos $\calG$), when the 
radius reaches infinitesimal dimensions, the components of the curvature 
become infinitesimal, and the ``singularity'' can be avoided. The model is not 
quite satisfactory since it does not show how the components of curvature from 
extremely large (almost ``infinite'') suddenly become infinitesimal. To gain 
an insight into this process, we change, in section 6, to another topos that 
is subtle enough to enable us to see what happens when a function diverges. 
This is the so-called Basel topos $\calB $, generated by a subtler Grothendieck 
topology on $\LL$ than that in the case of $\calG $. Unlike the topos 
$\calG$, it has invertible infinitesimals and infinitely large nonstandard 
natural numbers. They form a tool allowing us to manage divergences and, in 
some cases, ''go through a singularity''. Finally, in section 7, we make some 
general comments and a few remarks concerning  future perspectives.

\section{Space-Time Singularities}
In the present relativistic paradigm space-time is modeled by a pair $(M, g)$ 
where $M$ is a (four-dimensional) manifold and $g$ a Lorentz metric on $M$, 
with suitable differentiability conditions guaranteeing mathematical 
consistency and physically realistic interpretation. One should additionally 
assure that no regular points are removed from $M$, i. e. that space-time $(M, 
g)$ could not be extended with all differentiability conditions suitably 
preserved. If all these conditions are satisfied, and  
space-time is in some sense incomplete, this means that ``somewhere in it'' 
there is a singularity or, in other words, that this space-time is singular.

All these conditions should be made precise, and this is far from being 
straightforward. 
In the ``canonical'' singularity theorems \cite{HawkingEllis73} space-time 
incompleteness is usually understood as (timelike or null) geodesic 
incompleteness. However, from the physical point of view, it would be 
desirable to take into account not only geodesics but also all curves (of 
bounded acceleration\footnote{A curve of an unbounded acceleration could 
hardly be imagined to represent the history of a physically realistic 
object.}). Such a construction was proposed by Schmidt \cite{Schmidt71}, but 
it soon turned out that in physically realistic cases (such as closed Friedman 
and Schwarzschild solutions) it leads to a pathological behaviour. Several 
remedies were proposed to cure the situation \cite{Dodson78,HOPS07}. However, 
in the present work there is no need to go deeper into this side of the 
problem.

A space-time $(M', g')$ is said to be an extension of $(M, g)$ if there is an 
isometric embedding $\theta : M \rightarrow M'$ such that $\theta $ is onto 
and $\theta (M)$ is a proper subset of $M'$. A space-time is said to be 
extendible if it has an extension. Any space-time can be extended until it 
cannot be further extended, and if this is the case, such a space-time is 
called maximal. When we are hunting for singularities, we are looking for 
incomplete maximal space-times. The point is, however, that an extension is 
not, in general, unique. The most often obstacles are failures of various 
degrees of differentiability required to assure the existence of unique 
extensions (for other obstacles see \cite[p. 9]{Clarke}). 

Various degrees of differentiability are used in this context. Clarke 
\cite[pp. 63-64]{Clarke} argues that in order to extend a space-time through a 
suspected singularity one must go beyond the differentiability class assumed 
for this space-time. For instance, if one works, within a given space-time, 
with $C^k$-functions, for any $k$, then when attempting to make an extension, 
one must assume at least $C^{0, \alpha }$ differentiability\footnote{$C^{k, 
\alpha }$ differentiability means that $k$'th derivatives of a function must 
be H\"older continuous with exponent $\alpha $. A function $f$ is said to 
be H\"older continuous with exponent $\alpha $ if, for each compact 
domain $U$ of $f$, there is a constant $K$ such that $|f(x) - f(y)| < K |x 
- y]^{\alpha }$, for all $x, y \in U$.}.

When working with singular space-times, we want not only to know whether a 
singularity exists or not, but also how physics behaves on approaching the 
singularity. In this respect Riemann curvature tensor is more important than 
Lorentz metric. Unbounded metric could be a purely coordinate dependent effect 
whereas unbounded curvature tensor signifies unbounded tidal forces. This 
raises the problem of various degrees of differentiability of curvature tensor 
components. However, one should proceed here carefully since an unbounded 
curvature alone, does not indicate the existence of a singularity.

As we can see, in the singularity problem many things depend on the 
differentiability class of mathematical objects, that is to say on how these 
objects behave in a ``very small neighbourhood''. If we want to radically push 
forward the singularity problem, we should look for a method of dealing with 
space-time on the smallest possible scale. This is exactly what the so-called 
Synthetic Differential Geometry (SDG) is about.

\section{The SDG Strategy}
We begin with enriching the usual real line $\R $ by assuming the existence of 
infinitesimals, such as
\[
D:=\{x \in \R | x^2 =0\},
\]
$x \in D$ is so small (but not necessarily equal to zero) that $x^2 =  0$. 
The real line, enriched in this way, will be denoted by $R$. Let us notice the 
following 
consequence of the existence of $D$. Suppose we want to compute the derivative 
of the function $f(x) = x^2$ at $x = c$. For $d \in D$ we have
\[
f(c+d) = (c+d)^2 = c^2+2cd + d^2 = c^2 + (2c)d.
\]
The linear part of the latter expression can be safely identified with the 
derivative of $f(x)$ at $c$, $f'(c) = 2c$. We change this example into the 
rule by assuming

\textbf{Axiom 1.} For any $g: D \rightarrow R$, there exists a unique $b \in 
R$ such that
\[
\forall d \in D, g(d) = g(0) + d \cdot b.
\]
Let us notice that Axiom 1 states that the graph of $g$ coincides with a 
fragment of the straight line through $(0, g(0))$ and the slope $b$. And we can 
define the derivative of any function $f: R \rightarrow R$ at $c$ to be $f'(c) 
= b$. The important consequence of Axiom 1 is that \textbf{every function has 
a derivative}. In fact, in SDG all functions are differentiable (our 
presentation  
here follows \cite{Pizza}).

If this is so, all our problems with differentiability and extendibility in 
dealing with space-time singularities are eliminated with one stroke. But a 
high price is to be paid for such a step.

Let us consider the function 
\[
g(d) =
\left\{
\begin{array}{cc}
1 & \mbox{if $d \neq 0$} \\
0 & \mbox{if $d = 0$}
\end{array}
\right.
\]
From Axiom 1 we have: $D \neq \{0 \}$. Therefore, by the law of excluded 
middle, there exists $d_0 \neq 0$ in $D$ and, on the strength of Axiom 1,
\[
g(d_0) = g(0) + d \cdot b
\]
which lead to $1 = g(d_0) = d \cdot b$, and after squaring we get $1 = 0$.

There is only one strategy to save our ``radical way'' of dealing with 
space-time 
singularities -- to block the law of excluded middle. And indeed the SDG is 
founded on this strategy. It works on the basis of weakening classical logic 
to the inuitionistic logic (in which the law of excluded middle is not valid). 
Of course, one cannot change logic at will. This would lead to the complete 
mental anarchy. A modified logic requires a correct structural environment 
that would not only justify but also enforce the correct modification of 
logic. In the case of SDG this environment is provided by suitable topoi.

There are two approaches to constructing SDG: one can first formulate it in 
the form of an axiomatic system and then look for suitable topoi as its 
models \cite{Kock09}, or one first studies special cases of some interesting 
topoi and then generalises obtained results to the form of suitable axioms 
\cite{MoerRey}. Since we are only exploring possibilities of this method as 
far as the singularity problem is concerned, we work with a topos that seems 
best suited to this end, the one that generalises the category $\MM $ of 
manifolds (and smooth maps) in such a way that it includes manifolds with (at 
least some of) singularities. Only in section 6 we shall explore possibilities 
of another topos.

\section{Categorical Environment of Singularities}
Let $A$ be an $\R $-algebra, i.e. a commutative unitary algebra equipped with 
a homomorphism $\R \rightarrow A$. A $C^{\infty }$-ring is an $\R $-ring such 
that, for each smooth map $f: \R^n \rightarrow \R^m$ there is a smooth map 
$A(f): A^n \rightarrow A^m$ which preserves compositions, identities and 
projections. A homomorphism of $C^{\infty }$-rings ($C^{\infty } 
$-homomorphism) is a ring homomorphism $\phi : A \rightarrow B$ such that, for 
each smooth $f: \R^n \rightarrow \R^m$, one has
\[
\phi^m \circ A(f) = B(f) \circ \phi^n
\]
where $\phi^n: A^n \rightarrow B^n,\, \phi^m: A^m \rightarrow B^m$.

The crucial example of a $C^{\infty}$-ring is $\Cinf (M)$ where $M$ is a 
smooth manifold\footnote{In the following, we consider manifolds with a 
countable basis; in particular they are paracompact and can be embedded in a 
closed subspace of $\R^n$ for some $n$.}. Moreover, it can be shown \cite[p. 
24]{MoerRey} that if $M$ is a manifold, $\Cinf (M)$ is a finitely presented 
\Cinf -ring. Let us remind that a \Cinf -ring $A$ is finitely presented if it 
is isomorphic to a \Cinf -ring of the form $\Cinf(\R^m)/I$ where $I$ is a 
finitely generated ideal.

We are now ready to define the category of loci, our ``first approximation'' 
to determine the correct structural environment for the proposed categorical 
approach to the singularity problem. The category of loci, denoted by $\LL $, 
is the opposite (dual) category of the category of finitely presented $\Cinf 
$-rings as objects and $\Cinf $-homomorphisms as morphisms. Thus the objects 
are the same as in the category of finitely presented $\Cinf $-rings (for 
formal clarity we will distinguish an object $A$ and its formal dual $lA$) 
with arrows just reversed.

The category of loci contains the category of manifolds. More precisely, there 
is a functor $s: \MM \rightarrow \LL $ given by $s(M) = l\Cinf (M)$, where $M 
\in \MM $, which is full and faithful \cite[p. 60]{MoerRey}. It turns out that 
manifolds, put in the new environment, reveal new properties which were 
switched off in the old environment (in the category $\MM $).

To see this, let us first denote $l\Cinf (\R )$ by $R$, and analogously 
$l\Cinf (\R^n )$ by $R^n$. Let us also define the germ of $R$ at a point $p$ 
as
\[
\D_p = \bigcap\{s(U)| p \in U \; \mathrm{open \; in}\; \R \} \cong l\Cinf_p(\R 
),
\]
 and analogously for $p \in \R^n$,
\[
\D_p = \bigcap_{p \in U}s(U) \cong l\Cinf_p(\R^n).
\]
Let $lA $ be any locus with $A = \Cinf (\R^n)/I$, and let $p$ be a point of 
$lA$, i.e. a map $1 \rightarrow lA$ in  $\LL $, where $1 = R^0 = l\Cinf(\R^0)$ 
is the one point locus. Then $\D_p \cap lA$, where $\D_p \subset R^n$, is, by 
definition the germ of $lA$ at $p$. By applying this to a manifold $M \in \MM 
$, we obtain
\[
\D_p \cap s(M) \cong l\Cinf_p(M) = \bigcap \{s(U)| p \in U \; \mathrm{open \, 
in} \; M\},
\]
the germ of $M$ (strictly speaking of $s(M)$) at $p$. And the germ of a 
function $M \rightarrow \R $ at $p$ is now simply the restriction of $f$ to 
the germ of $M$ at $p$.

Here we have a comment on this important result: ``Thus, whereas the usual 
category of manifolds $\MM $ is too small to contain spaces which are germs of 
manifolds at some point, these `very small submanifolds' do exists in $\LL $'' 
\cite[p. 64]{MoerRey}. It is rather obvious that taking into account the 
existence of these ``very small submanifolds'' should change the strategy of 
dealing with singularities.

And what about higher classes of differentiability? Let us define 
\[
D_k(n) = l(\Cinf_0(\R ^n)/(x^{k+1})) = \{x \in R^n|x^{\alpha } = 0, \forall 
\alpha \mathrm{\; such \; that} \, |\alpha | = k+1\}.
\]
We immediately have
\[
D(n) := D_1(n) \subset D_2(n) \subset D_3(n) \subset \ldots ;
\]
All these infinitesimals are contained in $\D^n_0 \subset R^n$ where $\D_0 = 
\bigcap_{n \in \NN } s(-\frac{1}{n}, \frac{1}{n}) \subset R$.

Let $f: R^n \rightarrow R$ be a map in $\LL $. It corresponds to a smooth 
function $F: \R^n \rightarrow \R $, and the $k$-jet $j_n^k(F)$ of $F$, i.e. 
the equivalence class of $F|_0$ modulo $m^{k+1}$, is just the restriction 
$f|_{D_k(n)}$.

Of course, all these infinitesimals can be defined not only at $0$ of $R^n$ 
but also at any point $p$ of $s(M)$.

Smoothness of manifolds in $\LL $ is ``exact'' to such an extent that any 
manifold in $\LL $ has the tangent space when restricted to an infinitesimal
neighbourhood. To be more precise, let $M \in \MM $ and let $TM$ be the total 
space of the tangent bundle. Then we have
\[
s(M)^D \cong s(TM)
\]
with the projection
\[
ev_0: s(M)^D \rightarrow s(M).
\]
It is not a surprise that the tangent vector at a point $p \in M$ is just a 
point in $\LL $, namely $v: 1 \rightarrow s(M)^D$, such that $ev_0 \circ v = 
p$ \cite[pp.67-68]{MoerRey}.

In spite of all its advantages, the category $\LL $ of loci is not a good 
environment for doing differential geometry. In general, it has no 
exponentials of the form $l(A)^{l(B)}$, it is not Cartesian 
closed\footnote{Only duals of Weil algebras have exponentials in $\LL $; 
loosely speaking, only if $l(B)$ in $l(A)^{l(B)}$ is ``sufficiently small''.}, 
and consequently not even a topos. This makes it difficult to adapt techniques 
known from the category SET of sets and maps between sets to use them in $\LL 
$. 
We improve the situation in the following way.

Let us form the category $\mathrm{SET}^{\LL^{op}}$ the objects of which are 
functors from the category $\LL^{op}$, opposite with respect to $\LL $, to 
SET, and morphisms are natural transformations between these functors. The 
category $\mathrm{SET}^{\LL^{op}}$ has the presheaf structure. It contains the 
category $\LL $, the inclusion being given by the Yoneda embedding
\[
Y: \LL \hookrightarrow \mathrm{SET}^{\LL^{op}},
\]
\[
Y(lA) = \mathrm{Hom}_{\LL }(- , lA).
\]
It preserves all exponentiations $\LL $ possesses and is Cartesian closed. The 
category $\MM $ of manifolds sits (fully and faithfully) in 
$\mathrm{SET}^{\LL^{op}}$. To see this it is enough to make the composition $Y 
\circ s$. However, we are still not quite happy with this categorical 
environment for manifolds. The category $\mathrm{SET}^{\LL^{op}}$ has some 
``unwanted'' properties (for instance the enriched real line $R$ is in it 
non-Archimedean\footnote{$R$ is Archimedean if for every $x \in R$ there 
exists $n \in \NN $ such that $x < n$.}). To straighten them out we must 
impose the correct topology. This is done by converting presheaf into a sheaf. 
There are a few ways of doing that. The following way seems best suited for 
our purposes. Roughly speaking, we restrict the category 
$\mathrm{SET}^{\LL^{op}}$ to those functors ``which believe that open covers 
of $\MM $ are covers in $\mathrm{SET}^{\LL^{op}}$'' \cite[p. 97]{MoerRey}. 
More precisely, we consider a subcategory, let us call it $\GG $, of 
${\LL}$, the objects of which are duals $l(\Cinf (\R ^n)/I)$ 
of finitely generated $\Cinf $-rings, $I$ being a germ determined ideal. We 
define a suitable Grothendieck topology on $\GG $.\footnote{Going into details 
would blow up the limits of the present paper and, anyway, they are not 
essential to understand our main argument. The interested reader could consult 
\cite[pp. 98-101]{MoerRey}.} This topology (being subcanonical) gives rise to 
the 
category of sheaves on $\GG$ and we obtain a (Grothendieck) topos, 
denoted by $\calG $ and called the topos of germ determined ideals.
In what follows, we regard it as providing the correct categorical environment 
for our 
analysis.

The standard theory of general relativity is formulated within the SET 
category. It is clear that we cannot just jump to another category. We should 
rather carefully establish the relationship between categories SET and $\calG 
$, and see how in this relationship the category $\MM $ is situated. The 
general setting is given by the following diagram
\[
\MM \stackrel{s}{\hookrightarrow} \calG \stackrel{\Gamma }{\rightarrow} 
\mathrm{SET}
\]
As we remember, the category $\MM $ sits in $\mathrm{SET}^{\LL^{op }}$ fully 
and faithfully, and $\calG $ is just a suitable restriction of the 
latter\footnote{The composition $Y \circ s$ is customarily also denoted by 
$s$.}. The functor $\Gamma $, called the global sections functor, is defined by 
$\Gamma(F) = F(1)$ where $1$ is the terminal object of the category. This 
functor has the left adjoint functor $ \Delta: \mathrm{SET} \rightarrow \calG 
$,  called  the ``constant set'' functor, and $\Delta (S)$, $S \in 
\mathrm{SET}$, is just an ordinary set\footnote{Formally, $\Delta (S)(lA) = S$ 
for all $lA \in \LL^{ob}$.}.

\section{Singularities and Curvature}
In  this section, we present a simplified (toy) model illustrating the 
interaction between singularities and curvature. Strict results would require 
more detailed analysis which is under way. The usual $\Cinf $-manifold concept 
has its SDG generalisation as the formal manifold concept. It is just a smooth 
manifold equipped with an infinitesimal extension\footnote{A smooth 
$n$-dimensional manifold $M$, as an object in a suitable category, is a formal 
manifold if it is equipped with an open cover $\phi_i : \{U_i \rightarrow M\}$ 
by formally etal\'e monomorphisms, where $U_i$ are model objects, i.e. formal 
etal\'e subobjects in $R^n$ \cite{Kock80}.}. In the following, we consider 
formal manifolds in the category $\calG $. In \cite{MHJK16} we have 
constructed an infinitesimal version of $n$-dimensional formal manifold, i.e. 
a formal manifold $M$ with local maps of the form $\{ (D_{\infty})^n_i 
\rightarrowtail M|i\in I\}$ where $ 
D_{\infty}^n=\bigcup_{k=1}^{\infty}D_k(n)$, and we have shown that the 
curvature tensor ${\cal R}$ of any locally $D^n_{\infty}$-formal manifold, 
assumes only infinitesimal values in the object $D_k(m)$ for some $k\in 
\mathbb{N}$ and $m>n, m, n \in \mathbb{N}$. This result has important 
consequences for the singularity problem. We illustrate them with the help of 
the following simplified model.

Let us consider a model for an evolving universe given by
\[
S^3 \times \mathbb{R}\subset \mathbb{R}^4 
\]
where, in analogy with the closed Friedman-Lema\^{\i}tre world model,  
$\mathbb{R}$ can be interpreted as a cosmic time, and $S^3$ as a 3-dimensional 
instantaneous time section. Let us suppose that the smooth evolution 
$S^3\times \mathbb{R}$, described below, respects the standard smoothness of 
$\mathbb{R}^4$, i. e. the unique smooth structure on $\mathbb{R}^4$ in which 
the 
product $\R \times \R \times \R \times \R $ is a smooth product.

Let us farther assume that the diameter $\rho_{S^3}$ of $S^3$ shrinks to zero 
size (i.e., to a point in $\mathbb{R}^4$) which we call ``singularity'' and 
situate it at, say, $x_0=0, \, x_0 \in \mathbb{R}$. Topologically, we have a 
cone over $S^3$ with the vertex at the singularity. We should remember that it 
is a simple cone singularity rather than a curvature singularity met in 
standard cosmological models. If we delete the vertex together with its open 
neighbourhood, the cone is a standard smooth open submanifold of 
$\mathbb{R}^4$. If we leave the vertex in the picture, we obtain a toy model 
of a contracting universe ending its evolution in the singularity. 

This is how the evolution looks like in the category SET (which the standard 
relativistic cosmology tacitly assumes). How this evolution can be described 
when regarded as happening inside the topos $\calG $? We have
\[
(S_{\calG}^3 \times_{\calG} R) \hookrightarrow R^4.
\]
Let us notice that $S_{\calG}^3 \neq S^3$ because now $S_{\calG}^3$ is 
enriched by infinitesimals.  Since in $\calG $ infinitesimals appear, we can 
call it a fundamental or micro level, whereas SET will represent the macro 
level. 

$S_{\calG}^3 \times_{\calG } R$ becomes a locally $D^4_{\infty 
}$-infinitesimal formal manifold. If $S_{\calG}^3$ contracts, its 3-curvature 
grows, but when its radius reaches infinitesimal size, the components of the 
curvature become infinitesimal (if not zero). In this way, the conic 
singularity is avoided (and the evolution can be prolonged beyond $0$ of $R$).

We have an interesting result: from the macro point of view (SET perspective), 
the world has the initial singularity; from the micro point of view ($\calG $ 
perspective) the evolution is smooth (no singularity).

We have also the functor $ \Gamma : \calG \rightarrow \mathrm{SET} $. It tells 
us how the macro-image of the world emerges out of the micro-level. $\Gamma $ 
acts in this way that it evaluates the functor $F: \LL^{op} \rightarrow 
\mathrm{SET}$
(in $\calG $) at one point locus $1 = l(\Cinf (\R )/x)$, $\Gamma (F) = F(1)$. 
Owing to the fact that $\Gamma $ has a left adjoint $\Delta $, it enjoys nice 
properties from the point of view of category theory (it preserves inverse 
limits \cite[pp. 104-105]{MoerRey}).

\section{Passing through some Divergences of Curvature in a Smooth Topos - An 
Example}
In spite of its attractiveness, the simplified model presented in the 
preceding section has one serious disadvantage. It does not tell us how the 
curvature tending to infinity suddenly becomes infinitesimal (or whether there 
is no obstacle preventing such an outcome). In the present section, we address 
this problem, but to do so we must choose another topos whose inner 
environment is sensitive enough to see what happens on approaching the 
singularity. The topos we will work with is called the Basel topos, denoted by 
$\calB $. It is constructed in the following way. On the category $\LL $ of 
loci we define a suitable Grothendieck topology\footnote{Consisting of finite 
open covers and projections \cite[pp. 285-286]{MoerRey}.}. $\LL $ with this 
topology is called the site $\mathbb{B}$, and the Basel topos $\calB $ is the 
topos of sheaves on $\mathbb{B}$. The Basel topos has much more complex 
structure than the topos $\calG $, but many properties of $\calG $ are 
incorporated in $\calB $. In particular, 
the category $\MM $ of manifolds also sits fully and faithfully in $\calB $. 
However, this richer structure of $\calB $ is indispensable to deal with such 
notions as convergence or ``going to a limit''.

The space of infinitesimals in $\calB $ is given by
\[
\D = \{x \in R| \forall n \in \NN \; -\frac{1}{n+1} < x < \frac{1}{n+1}\}.
\]
It has two subspaces
\[
\II = \{x \in \D| x \; \mathrm{is \; invertible}\},
\]
and
\[\Delta = \{x \in R| \mathrm{x \; is \; not \; invertible}\}.
\]
In $\calB $ we have $D\subset \Delta \subsetneq \D  \subset R$ and $\II \neq \emptyset$. 
This is the consequence of the existence in ${\cal B}$ the natural numbers 
$N$, called also the smooth natural numbers, which are end-extension of the 
standard $\mathbb{N}$, i.e. $\mathbb{N}\subsetneq N$. Then $\mathbb{I}=\{ x\in 
R|\exists n\in N\setminus \mathbb{N}\, (x<\frac{1}{n+1}\vee 
x>\frac{-1}{n+1}) \}$ and $\Delta =\{x\in R|\forall n\in N (-\frac{1}{n+1}<x<\frac{1}{n+1})  \}$. Since $\mathbb{N}\subsetneq N$ and $\D_0 = \bigcap_{n \in \NN } 
s(-\frac{1}{n}, \frac{1}{n}) \subset R$, it is 
now clear that also $\Delta \subsetneq \D_0$. 

Let us now consider a real function $f:\mathbb{R}\to \mathbb{R}$ in SET which 
has divergent behaviour near $0\in \mathbb{R}$, i.e. $\underset{x\to 0^+}{\lim 
}f(x)=+\infty$.  In our example of a 3-sphere with its radius shrinking to 
zero at $0\in \mathbb{R}$, the scalar curvature of the sphere is simply given 
by $f(x)=\frac{1}{x}$. In general, we can have a function of arbitrary fast 
divergence. We will show that the smooth topos ${\cal B}$ gives us a tool 
allowing 
for smoothly overpassing such divergences.

Let us first approximate the divergent behaviour of $f(x)$ near $0$ in SET by 
the sequence $(g_k)$ of smooth functions on the whole of $\mathbb{R}$, which 
are convergent to the distribution $\delta$. Of course, the convergence is 
understood in the distributional sense. Our minimal requirement is that each 
$g_k, k=0,1,2,...$, continuously prolongs the preceding function $g_{k-1}(x), 
k=1,2,...$; and we assume that $g_0$ prolongs $f(x)$ starting from some 
$\tilde{x}>0$. Let us consider a $\delta$-like sequence of functions, 
$(\delta_{\epsilon}(x))$.\footnote{$\underset{\epsilon\to 0}{\lim} 
\delta_{\epsilon}(x)\overset{{\rm distrib.}}{=}\delta$ $(\underset{k\to 
\infty}{\lim} 
\delta_{k}(x)\overset{{\rm distrib.}}{=}\delta)$ $\iff$ for all test functions 
$\phi(x)\in S$ it holds $\underset{\epsilon\to 0}{\lim 
}\int_{\mathbb{R}}\phi(x)\delta_{\epsilon}(x)dx=\phi(0)$ $(\underset{k\to 
\infty}{\lim }\int_{\mathbb{R}}\phi(x)\delta_{k}(x)dx=\phi(0))$.} We build the 
sequence $(g_k)$ with the help of $(\delta_{\epsilon}(x))$.

However, if we want to prolong the function $f(x)$ by a member 
$\delta_{\epsilon}(x)$, say, of the sequence, and then by another and so on, 
it can happen that $f(x)$ and $\delta_{\epsilon}(x)$ do not meet for any 
$x>0$ and $\epsilon>0$, so that the prolongation would not be continuous. 
In such a case we can always take $A\cdot \delta_{\epsilon}(x)$ with $A\in 
\mathbb{R}$ such that for some $x,\epsilon \in \mathbb{R}_{>0}$, 
$f(x)>\delta_{\epsilon}(x)>0$ and $A\geq 
\frac{f(x)}{\delta_{\epsilon}(x)}$. Now, $A\cdot \delta_{\epsilon}(x)$ meets 
$f(x)$, and the sequence $(A\cdot \delta_{\epsilon}(x))$ is still a 
$\delta$-like sequence converging distributionally to the distribution $A\cdot 
\delta$.
 
Having this in mind, let us define the family of functions (subsequent 
prolongations of $f(x)$) with $A$ depending on $k$
\begin{equation}\label{e1} g_{k}(x)= \begin{cases} 
      f(x) & x\in [\eta_0,+\infty) \\
      A(k)\cdot \delta_{k}(x) & 0< x\leq \eta_k , A(0)\cdot 
\delta_{0}(\eta_0)=f(\eta_0)\\
            & A(l+1)\cdot 
\delta_{l+1}(\eta_l)=\delta_l(\eta_l),l=1,2,..,k-1.\\
            & \eta_{l+1}\leq \eta_{l}, l=0,1,2,.,k-1.
   \end{cases}
\end{equation}
One has $\underset{k\to \infty}{\lim}\eta_k=0, \underset{k\to \infty}{\lim} 
g_{k}(x)= +\infty$
and $g_k(x)=f(x), x\in [\eta_0,+\infty)$. In fact, all $A(k), k>0$, can be 
chosen equal to $1$, since $\delta_k(x)=\delta_{k+1}(x)$ for some $x\in 
(0,\tilde{\eta}_k ), k=1,2,...$, and $\frac{\delta_{k+1}(x)}{\delta_k(x)}=1$ 
at this $x$. This follows from the behaviour of continuous functions in the 
$\delta$-like sequence. By choosing this $x$ as $\eta_k$, we have the new 
interval $(0,\eta_k]$ where at $x=\eta_k$ both functions, $\delta_k(x)$ and 
$\delta_{k+1}(x)$, assume the same value. In this way, $g_k(x), x\in 
(x_0,\infty)$, 
is a continuous function for any $x_0>0$ (in fact it can be smoothed out). 
Increasing $k$ means decreasing $x_0$. However, no continuous function, for 
all $ x\in [0,\infty)$, can be a limit of $g_k(x)$ when $k \rightarrow \infty 
$. The correct limit has to be distributional. 
To describe it, let us define ``the $\delta$-approximation of infinity'' by 
the prolongations of $f(x)$, i.e. the sequence of functions

\begin{equation}\label{e2} \tilde{g}_{k}(x)= \begin{cases} 
      g_k(x) & x\in (\eta_k,+\infty) \\
       A\delta_k(x) & x\in (-\infty, \eta_k]
   \end{cases}
\end{equation}
The distributional limit $\underset{k\to \infty}{{\lim}_{\rm 
distr.}}\tilde{g}_k(x)$ is the pair: the regular distribution represented by 
the function $g_k, k\to \infty$ (the subsequent prolongations) for $x\in 
(0,\infty)$, and the $A\delta$-distribution, i.e. the distributional limit of 
the $\delta$-like sequence $\delta_k(x), x\in (-\infty, \eta_k], \eta_k 
\underset{k \to \infty}{\to} 0$.  

The construction of the sequence $(g_k(x))$ of the subsequent prolongations of 
$f(x)$ gives the 
translation of the divergence of $f(x)$ into the $\delta$-like divergence of 
$g_k(x)$. We call it the $\delta$-like divergence approximating the divergence 
of a function 
$f$. In SET there is no way to pass smoothly or continuously over such a 
distributional divergence.  
However, in ${\cal B}$ the $\delta$-like divergence disappears and is 
replaced by the smooth internal evolution over $R$. This is accomplished by 
the following theorems. 

Let $F_n$ be the internal space in $\cal B$ of test functions in dimension $n$ 
\cite[p. 322]{MoerRey}. 

\begin{Theorem}[\cite{MoerRey} Theorem 3.6, p. 324]\label{Th1}
Every distribution $\mu$ on $R^n$ in $\cal B$ can be represented by a function 
(predistribution) $\mu_0:R^n_{acc}\to R$ such that for all $f\in F_n$, 
$\mu(f)=\int f(x)\mu_0(x)\,dx$. Here $R_{acc} = \{x \in R|\, \exists n \in \NN 
(-n < x < n)\}$ (called accessible reals).
\end{Theorem} 

The following theorem is based on the $\delta$-like sequence 
$\delta_{\epsilon}(x)=\frac{1}{\pi 
\epsilon}\frac{\sin{x/\epsilon}}{x/\epsilon}=\frac{1}{\pi}\frac{\sin{x/\epsilon}}{x}$ 

which, in terms of $\epsilon\simeq 1/n$ reads 
$f_n(x)=\frac{1}{\pi}\frac{\sin{nx}}{x}$.
\begin{Theorem}[\cite{MoerRey} Theorem3.7.1, p. 325]\label{th2}
For every $n\in N\setminus \mathbb{N}$ ($n>\infty$) and every $f\in F_1$
\[ \int\frac{\sin{nx}}{\pi x}f(x)dx \simeq f(0).\]
\end{Theorem}
Thus we can always represent a $\delta$-distribution by the function 
$\delta_0=\mu_0$ making $\delta$ the regular distribution in $\cal B$, where 
$N$ is the object of smooth natural numbers. This has tremendous consequences. 
The following theorem makes the direct use of a regular function, internal in 
$\cal B$, on the level $n>\infty$, replacing the $\delta$-distribution in 
$g(x)$ of (\ref{e2}) in SET.
\begin{Theorem}\label{th3}
The $\delta$-like divergence in SET approximating the divergence of a function 
$f$, $f\in 
C^0(\mathbb{R}_{>0})$, $\underset{x\to 0^+}{\lim}f(x)=+\infty$, is 
realized in ${\cal B}$ by the internal $s$-evolution, $g_{\cal B}:R\to R$ 
smooth at $0$, in $\cal B$ 
($s$ -- after ``smooth'', i.e. evolution in terms of 
smooth natural numbers). $\square $
\end{Theorem}
One easily extends this result, based on Theorem \ref{Th1}, to the 
divergences generated by $f:\RR^n\to\RR, \; n>1$.
We see that in order to make a geometrical use of noninvertible infinitesimals 
from $D^n_{\infty}$, we must first ``reach them'' with the help of non-zero 
invertible infinitesimals from $\mathbb{I}$. Happily enough they exists in 
$\cal B$, and their inversions, although infinite, help to define new 
$s$-finiteness in $\cal B$. 

\section{Final Remarks}
As we have remarked in the Introduction, this is only an introductory work, 
the aim of which is to convince the reader that the topic is worthwhile to be 
pursuit, rather than to present concrete results. New tools provided by SDG, 
non available in the standard approach, give us an opportunity to investigate 
what happens on ``infinitesimally small neighbourhoods'' and when various 
processes ``go to infinity''. The first candidate in this line of research, 
taken up in the present 
work, is the singularity problem. As the next step one could investigate 
various kinds of singularities (strong curvature singularities, quasiregular 
singularities, etc.) and singularities in various solutions to Einstein's 
equations (see \cite{Grin96}). Other open fields are: geodesic incompleteness, 
various space-time extensions, and space-time boundaries of different kinds; 
everything as seen in the light of SDG possibilities.

Is seems worthwhile to try also categorical and SDG methods in the search for 
quantum gravity theory (see \cite{Duplij}), especially that the very intensive 
search with the help of other methods has so far given rather modest results.

A remarkable circumstance is that the categorical approach not only can 
contribute at solving, or at least elucidating, some existing problems, but 
also is able to uncover new facts and regularities. For instance, interaction 
between SET and another topos (such as $\calB $), via some functors, can 
produce in $\R^4$ an exotic smooth structure, and since no such structure can 
have vanishing Riemann tensor, this effect can lead to interesting results 
(see \cite{TAMJK2013,MHJK16}).

However, there is a price one has to pay for all these advantages: when 
changing from SET to another topos, one must be ready to switch from classical 
to intuitionistic logic. This is a radical step. Our brain has evolved 
through a long interaction with its macroscopic environment, the logical 
structure of which is shaped by the internal logic of the topos SET, i.e. 
classical logic. However, the logic of the entire universe, on all its levels, 
is not obliged to conform to the logic of our brains.

\end{document}